\documentclass{article}

\usepackage{arxiv}

\usepackage[utf8]{inputenc} % allow utf-8 input
\usepackage[T1]{fontenc}    % use 8-bit T1 fonts
\usepackage{hyperref}       % hyperlinks
\usepackage{url}            % simple URL typesetting
\usepackage{booktabs}       % professional-quality tables
\usepackage{amsfonts}       % blackboard math symbols
\usepackage{nicefrac}       % compact symbols for 1/2, etc.
\usepackage{microtype}      % microtypography
\usepackage{lipsum}		% Can be removed after putting your text content
\usepackage{graphicx}
\usepackage{natbib}
\usepackage{doi}
\usepackage{bm}
\usepackage[varvw]{newtxmath}       % selects Times Roman as basic font
\usepackage{amsmath}
\mathchardef\mhyphen="2D % Define a "math hyphen"
\usepackage{booktabs}
\usepackage{multirow}
\usepackage{comment}

\title{Recent Advances in Uncertainty Quantification Methods for Engineering Problems}

\author{ {Dinesh ~Kumar } \\
	Department of Mechanical Engineering\\
	University of Bristol\\
	Bristol BS8 1TR, UK \\
	\And
         {Farid ~Ahmed} \\
	Department of Nuclear Engineering\\
	North Carolina State University \\
           2500 Stinson Drive Raleigh, NC, USA
\And
         {Shoaib ~Usman } \\
	Nuclear Engineering and Radiation Science\\
	Missouri University of Science and Technology \\
	Rolla, MO 65409, USA \\
 \And
         {Ayodeji ~Alajo } \\
	Nuclear Engineering and Radiation Science\\
	Missouri University of Science and Technology\\
	Rolla, MO 65409, USA \\
 \And
      {Syed ~Alam} \\
	Nuclear Engineering and Radiation Science\\
	Missouri University of Science and Technology\\
	Rolla, MO 65409, USA \\
 }

\renewcommand{\shorttitle}{\textit{AI Assurance: Towards Trustworthy, Explainable, Safe, and Ethical AI } }

\hypersetup{
pdftitle={A template for the arxiv style},
pdfsubject={q-bio.NC, q-bio.QM},
pdfauthor={David S.~Hippocampus, Elias D.~Striatum},
pdfkeywords={First keyword, Second keyword, More},
}

\begin{document}
\maketitle

\begin{abstract}
In the last few decades, uncertainty quantification (UQ) methods have been used widely to ensure the robustness of engineering designs. This chapter aims to detail recent advances in popular uncertainty quantification methods used in engineering applications. This chapter describes the two most popular meta-modeling methods for uncertainty quantification suitable for engineering applications (Polynomial Chaos Method and Gaussian Process). Further, the UQ methods are applied to an engineering test problem under multiple uncertainties. The test problem considered here is a supersonic nozzle under operational uncertainties. For the deterministic solution, an open-source computational fluid dynamics (CFD)  solver SU2 is used. The UQ methods are developed in Matlab and are further combined with SU2 for the uncertainty and sensitivity estimates. The results are presented in terms of the mean and standard deviation of the output quantities.
\end{abstract}

% keywords can be removed

In industrial applications, during the operation of engineering devices, several properties and parameters of the components change with time. The material properties of its components vary continuously due to several operational factors. The failure point information of the week components in an industrial application is very useful. The safety of the engineering device under consideration should always be of utmost concern for the manufacturers while designing the device. Using statistical information, the designer can evaluate the safety margin or make the failure design margin smaller than other components so that the impact of the weak component can be minimized. With advancements in computer hardware and numerical algorithms, computational tools are used to design advanced and high-performance engineering components in almost all engineering fields. For example, aircraft, high-speed car manufacturers, sports, naval ship designers, etc., use computational fluid dynamics (CFD) to simulate fluids over the device. These computational tools are also used for thermal and structural analysis to simulate and detect faults, cracks, and failure of the devices by almost all industries.\\
 
 Uncertainties are an inherent part of the computing systems concerning real-world applications \cite{oberkampf2002verification,roy2011comprehensive,beyer2007robust,schueller_review_2008,wiener_chaos_1938,xiu_2002,smith2013uncertainty,kumar2021quantitative,kumar2020uncertainty,kabir2021uncertainty,kabir2021optimal,kabir2020uncertainty,kumar2021multi,kumar2020nuclear,kumar2019influence}. Two physical experiments can never produce the same results as several of the system parameters are not known properly and have uncertainties. When the same system is modeled using computational tools and mathematical equations, the input and system parameters are provided with constant values to predict the results without dealing with the uncertainties in the input parameters. Almost all aspects of engineering modeling and design are affected by these uncertainties. Engineers and researchers have always encountered issues related to uncertainties in terms of design reliability and robustness. By understanding sources of uncertainties and quantifying them, one can estimate confidence in the system outputs. In mathematical modeling, uncertainties are usually encountered in initial conditions, boundary conditions, material properties, weather conditions, and manufacturing tolerances. \\

Uncertainty Quantification (UQ) is the field of detecting, describing, quantifying, and managing uncertainties in computational designs of real-world systems. In UQ, the system response is estimated in a stochastic way by combining the deterministic solver with statistical tools. UQ methods are statistical tools to assess safety margins in the system responses when computer simulations are used to design an engineering device. UQ methods address the problems associated with incorporating system parameters variability and stochastic behavior into systems analyses. Computer simulations answer what happens when the system under consideration is subjected to a set of input parameters. However, UQ expands this question and answers what will happen when the system is subjected to a range of variability in the input parameters. UQ combines mathematics, statistics, and engineering. Generally speaking, UQ methods predominantly treat the system to be studied as a closed system (like a black box), and an extensive understanding of the system's inner functioning is not required. The UQ methods only need information about the input parameters of the model and model responses to estimate probabilistic model responses. In the recent past, uncertainty quantification and management are considered as significant elements in risk management (the system can fail or be damaged if it does not meet the design targets) of industrial designs \cite{oberkampf2002verification,hirsch2018uncertainty}.\\ 
  
Due to the non-intrusive nature of UQ methods, these methods can be adopted easily by researchers and industries from a wide range of engineering, industrial and financial sectors to:

\begin{itemize}
\item Understand the uncertainties inherent in the system.
\item Predict system responses concerning uncertain inputs.
\item Quantify confidence in the system responses.
\item Find optimal responses concerning a wide range of inputs.
\item Reduce unexpected system failures.
\item Implement probabilistic modeling and design processes.
\item Predict parametric sensitivity on the model responses.
\end{itemize}

With increasing computational power and simulation techniques, it became possible to make accurate predictions of real-world systems. Now the challenges in engineering designs are moved toward predicting system behaviors with respect to uncertainties efficiently. Traditional UQ methods based on Monte Carlo \cite{hammersley2013monte,rubinstein2016simulation} usually need a large number of system evaluations. So these methods are restricted to simplified test cases and for research purposes only. Monte Carlo methods are sampling-based methods, and the convergence rate is very slow. In general, large samples (at least $10^4$) are needed to predict statistical quantities accurately. Alternatively, the literature proposes several sampling schemes such as Latin Hypercube, Sparse sampling, clustered sampling, and stratified sampling to accelerate the convergence. However, Monte Carlo methods have not gained massive popularity due to their expensive computational cost. For large-scale problems and real-world engineering applications, more recent methods based on machine learning approaches such as Polynomial Chaos Method (PCM) \cite{xiu2002wiener, najm2009uncertainty,hosder2007efficient,ghanem2017handbook}, Gaussian Process (or Kriging) \cite{quinonero2005unifying,bastos2009diagnostics}, Support Vector Machine (SVM) \cite{awad2015support,smola2004tutorial}, Polynomial-Chaos-Kriging (or PC-Kriging) \cite{schobi2015polynomial,wang2019multi} are proposed in literature and are applied to several diverse applications. Polynomial chaos and Gaussian process models are seen as leading approaches for stochastic and robustness analyses of very complex engineering applications. PC-Kriging is a result of combining polynomial chaos and Kriging methods. These approaches are discussed in detail and are applied to an engineering application in the following sections. \\

In order to understand the potential influence of input parameters on system outputs, the Sensitivity Analysis (SA) method is used \cite{saltelli2002sensitivity,Sudret2008Global}. Various methods for global sensitivity analysis, such as linear regression and variance-based analysis are discussed for sensitivity estimation. A standard method of estimating system responses is using Sobol' indices-based global sensitivity analyses. Various meta-modeling methods can compute Sobol indexes, including Monte Carlo, graphical models, Kriging, and support vector machine approaches. In recent years, surrogate models that calculate Sobol' indices have gained considerable attention. The first step in this approach is to construct a surrogate model using the design of experiments (DOE). Furthermore, this surrogate model is used to estimate many model responses in order to compute Sobol' indices. Computing model responses from surrogate models are also called data-driven approaches, as several combinations of input parameters are used to explore a wide range of input domains. \\

\section{Polynomial Chaos Method for UQ}\label{PCEsec}

In 1938, Wiener proposed the polynomial chaos method for dealing with Gaussian distributed uncertainties \cite{wiener_chaos_1938}. Xiu and Karniadakis \cite{xiu_2002} demonstrated the ability to use it with any probability distribution in a detailed analysis. In the last few years, the generalized method has been used in a variety of engineering applications, including computational fluid dynamics, heat transfer, nuclear reactor design, and structural analysis \cite{kumar2020efficient}. Because adding uncertainties increases the computation required to quantify them, early applications dealt with a limited number of uncertainties. With increasing uncertainties, the number of simulations required to quantify the uncertainty grows exponentially using the polynomial chaos method. This is referred to as the dimensionality curse. Numerous improvements have been proposed in literature \cite{blatman2011adaptive,liu2020efficient,hosder2007efficient,kumar2016efficient,aremu2020machine,liu2019escaping} to cope with the curse of dimension and move the UQ process forward. Several researchers also proposed model reduction algorithms (based on principal component analysis) to accelerate the polynomial chaos method.\\

Numerous applications have used model reduction approaches. However, these approaches mainly were two steps processes and usually were applied in semi-intrusive ways. Thus, they were not very straightforward to use for engineering applications where the models can be used as a black-box. Several researchers proposed the idea of sparse sampling. Using sparse sampling schemes (such as Fejer, Clenshaw-Curtis, Conrod-Patterson), the number of simulations can be reduced to achieve the same accuracy as classical polynomial chaos. Blatman and Sudret proposed a theory of sparse polynomial chaos, based on least angle regression in their paper \cite{blatman2011adaptive,bourinet2018reliability}. Based on its principle, this method used a maximum number of polynomial order approximations for a given number of samples and a sparse polynomial chaos expansion (PCE) for a given system response \cite{kumar2020efficient,kumar2021quantitative,kumar2021multi,kumar2020uncertainty,kumar2016efficient}. Several other researchers also proposed the more or less similar idea of sparse polynomial chaos using different error minimizing schemes. Recently, numerous applications have seen the sparse polynomial chaos approach due to its straightforward usage and faster convergence capability. In this section, some fundamental concepts for the polynomial chaos approach are described \cite{ghanem2017handbook}. It is important addressing that the lead author developed this method and the descriptions have been reported in different studies \cite{kumar2020efficient,kumar2021quantitative,kumar2021multi,kumar2020uncertainty,kumar2016efficient} for a range of engineering applications.  \\

Based on a set of orthogonal polynomial basis functions, we can write a stochastic model response for a system under uncertainty as follows:

\begin{equation}
\label{Y1}
	\displaystyle Y=M(\bm \xi)= \sum_{\bm b \in \mathbb{N}^n} a_{\bm b}\psi_{\bm b}(\bm \xi)
\end{equation}

where $\psi_{\bm b}$  is an orthogonal polynomial for multidimensional dimensions, $ \bm b=b_1\dots b_n $ represents an index and the terms $ a_{\bm b} $ are called polynomial coefficients. A PCE is given by the Equation (\ref{Y1}). A system of $n$-dimensional input uncertainties is represented by $\bm \xi$ in the above equation. From a set of orthogonal one-dimensional polynomials, we construct the multi-dimensional polynomials $\psi_{\bm b}$ as follows  \cite{kumar2016efficient}:

\begin{equation}
\label{Y2}
	 \psi_{\bm b}(\bm \xi) = \psi_{b_1\dots b_n}(\bm \xi) = \prod_{i=1}^n \psi_{b_i}(\xi_i)
\end{equation}

where $ b_i $ is the order of the polynomial expansion for the random variable $ \xi_i $. \\

Extended polynomial expansions usually truncate to a finite number of terms because higher-order terms are not significant in the system response after a few terms. We truncate PCE into the following in order to achieve the following degree $ |\bm b |= \sum_{i=1}^nb_i $ within a given order $ p $ \cite{du2019efficient}:

\begin{equation}
\displaystyle Y\simeq M_p(\bm \xi)= \sum_{\bm b \in A^{p,n}} a_{\bm b}\psi_{\bm b}(\bm \xi),~A^{p,n}=\{\bm b \in \mathbb{N}^n : |\bm b| \leq p \}
\end{equation}

The total number of terms, $P$ (basis functions), equals $\frac{(n+p)!}{n!p!}$ when the number of input uncertainties is $n$ and the highest order of polynomial in PCE is $p$. Polynomial coefficients can be calculated based on the PCE order and solution samples (system responses using a deterministic solver as black box). One can compute and construct the PCE of a stochastic output. In the PCE, the first term (the zeroth-order term) represents the stochastic response's mean. In addition, one can also compute higher-order statistical moments numerically by using these polynomial coefficients. Computing polynomial coefficients can be done using numerical methods such as collocation and regression \cite{kumar2016efficient}. \\

Once we calculate polynomial coefficients, the mean $E(Y)$ and variance $V(Y) $ of the system output $Y$ can be computed easily as below:

\begin{equation}
E(Y) = a_0;  V(Y) = \sum_{i=1}^{P} a_i^2 \psi_i^2
\end{equation}

where the coefficients $a_i$ and $\psi_i$ are the same as they were defined earlier.

%%%%%%%%%%%%%%%%%%%%%%%%%%%%%%%%%%%%%%%%%%%%%%%%%%%%%%%%%%%%%%%%%%%%%%%%%%%%%%%%

\section{Gaussian Process or Kriging for UQ}\label{GPsec}

Kriging, also known as Gaussian process modeling, is a statistical method for approximating various functions and computer experiments using Gaussian processes. Kriging is also used as a surrogate model to establish a link between the inputs and outputs of expensive computational models \cite{zhang2016brownian}. The Gaussian process has been used for several machine learning applications related to regression and classification in the last few decades. Krige first developed kriging method for geostatistical applications in 1951. Further, it was used in metamodeling and data-driven modeling for numerous applications with noisy data. The model is known as kriging (after Krige in geostatistics). Using Gaussian processes, Tarantola and Valette designed a Bayesian formulation for inverse problems in geophysics \cite{tarantola1982inverse}. Based on the work of Williams, Neal, and Rasmussen, the model was proposed to solve regression problems in statistics \cite{o1978curve,Hinton1995wake,williams1996Gaussian} and gained popularity. \cite{Hinton1995wake,williams1996Gaussian,gibbs1997efficient} provides the Bayesian interpretation and detailed description of the model. Machine learning was introduced to Gaussian Process in the nineties. As a result of a detailed comparison by Rasmussen \cite{Hinton1995wake,williams1996Gaussian} of the GP with the most widely used models, the GP started becoming very popular. GP approaches outperformed other approaches in the vast majority of cases, he showed. Using the maximum-likelihood estimation method (MLE), the GP model's learning process involves tuning the covariance parameters to the data. In order to obtain the prediction and the degree of uncertainty associated with it, given a new input and conditioned on previous observations, one can easily calculate the mean and variance of the predictive distribution. By using the definition of conditional probabilities, we can easily obtain Gaussian distribution based on the GP assumption \cite{girard2004approximate}.\\

The output response of the model $M$, according to Kriging, is the realization of a Gaussian process. Kriging metamodel $M^K(x)$ of the true model $M(x)$ can be described as:

\begin{equation}
M^K(x)=\beta^Tf(x)+\sigma^2Z(x,\omega)
\end{equation}

Where $\beta^T f(x)$ is the mean of the Gaussian process, $\sigma^2$ is the variance of the process, and $Z(x, \omega)$ is a stationary Gaussian process with a zero mean and unit variance \cite{du2019efficient}. The underlying probability space  ($\omega$)  is defined in terms of a correlation function $R(x_1, x_2; \theta)$ that describes the correlation between two sample points in the output space $x_1$ and $x_2$, as well as the hyperparameters $\theta$.\\

If $y ={y_1, y_2, y_3, . . ., y_N}$ are the outputs of the true model $M(x)$ at sampling points $x= {x_1, x_2, x_3, . . ., x_N}$ , the model prediction $M^K(x)$ at a new point $x$ can be estimated using Kriging metamodeling. The gaussian process metamodeling prediction is based on the fact that the prediction $y’$ at the new point $x$ and the responses from the true model $y$ make a joint Gaussian distribution as:

\begin{equation}
\begin{Bmatrix}
y'\\ 
 y
\end{Bmatrix}
=\mathbb{N}_{N+1}\begin{pmatrix}
 & 
\begin{Bmatrix}
f^T(x )\beta \\ 
 F\beta
\end{Bmatrix}, \sigma^2
\begin{Bmatrix}
r(x)^Tr(x)\\ 
 R
\end{Bmatrix}
\end{pmatrix}
\end{equation}

In the above equation, $F$ is the observation matrix with entries $f_j(x_i)$ for $i =1, 2, 3. . ., N$ and $j=1, 2, 3, . . ., P$ where, $f_j(x_i)$ are arbitrary functions at observation points $x_i$ and $\beta$ are regression coefficients. The vector $r(x)$ is the cross correlations between the new point $x$ and the known points $x_i$ as:

\begin{equation}
\begin{Bmatrix}
r_1\\ 
 r_2\\
r_3\\
.\\
.\\
r_N
\end{Bmatrix}=
\begin{Bmatrix}
R(x,x_1, \theta) \\ 
R(x,x_2, \theta) \\ 
R(x,x_3, \theta) \\ 
.\\
.\\
R(x,x_N, \theta) \\ 
\end{Bmatrix}
\end{equation}
 
Where $R$ is the correlation matrix at the known points $x_i$ and can be written as:

\begin{equation}
R_{ij} =R(x_i, x_j; \theta)
\end{equation}

where $i, j =1,2,3,...N$. 

or 
\begin{equation}
R =
\begin{Bmatrix}
R(x_1,x_1, \theta) & R(x_1,x_2, \theta)  &. &.& .  & R(x_1,x_N, \theta) \\ 
R(x_2,x_1, \theta) &R(x_2,x_2, \theta) &. &.& . &R(x_2,x_N, \theta) \\ 
R(x_3,x_1, \theta) &R(x_3,x_2, \theta)& . &. &. &R(x_3,x_N, \theta) \\ 
. & . &. &. &. &.\\
. & . &. &. &. &.\\
R(x_N,x_1, \theta) &R(x_N,x_2, \theta)& .& . &. &R(x_N,x_N, \theta) \\ 
\end{Bmatrix}
\end{equation}

Using the conditional distribution properties of the multivariate normal, the mean and the variance of the predictor can be written as:

\begin{equation}
E\{y'|y\} = f^T \beta +r^T R^{-1}(y-F\beta)
\end{equation}

\begin{equation}
\sigma^2_y = \sigma^2(1-r^TR^{-1}r+u^T(F^TR^{-1}F)^{-1}u)
\end{equation}

where the regression coefficients $\beta$ and the term $u$ are defined as:
\begin{equation}
\beta = (F^TR^{-1}F)^{-1}F^TR^{-1}y
\end{equation}

\begin{equation}
u = F^TR^{-1}r - f
\end{equation}

\section{Polynomial Chaos Kriging for UQ}\label{GPsec}

Kriging interpolates local variations in the system response $Y$ as a function of the neighboring design points, whereas PCE closely approximates the regional behavior of $Y$ \cite{amini2021copula}. It is possible to obtain more accurate PC-Kriging metamodels by combining local and global approximation techniques. In PC-Kriging, there is an array of orthonormal polynomials that represent the trend and which are defined as follows:

\begin{equation}
M^{PCK}(x)=\sum_{\bm b \in A^{p,n}} a_{\bm b}\psi_{\bm b}(\bm \xi)+\sigma^2Z(x,\omega)
\end{equation}

Where $\psi_b$ are multivariate orthogonal polynomials concerning the input distributions and $a_b$ are the corresponding coefficients.\\

 \section{Uncertainty Quantification of a Supersonic Nozzle}

\begin{figure}[htbp]
 \centering
 \includegraphics[width=0.8\textwidth]{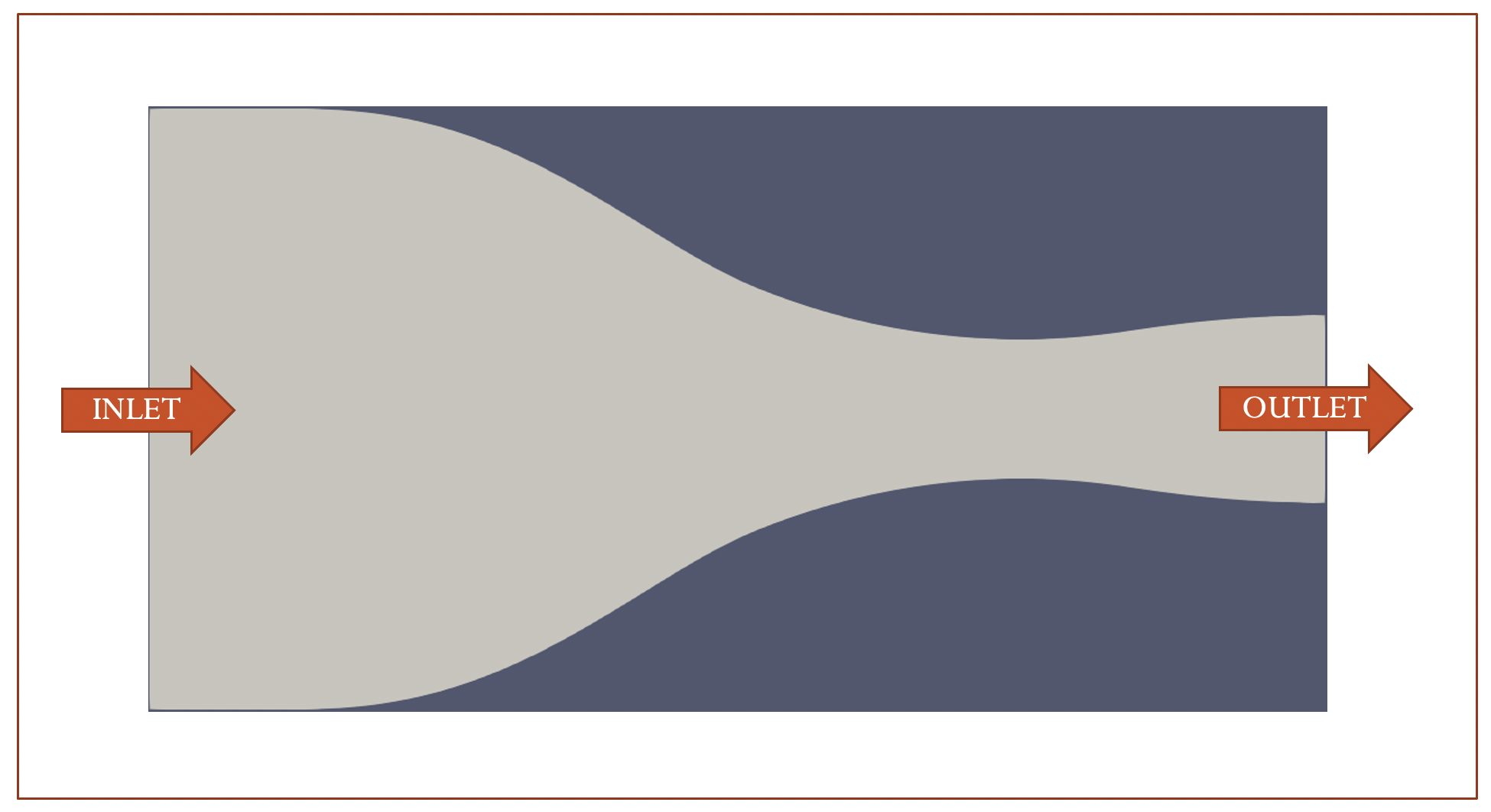}\\
 \caption{CD Nozzle}\label{PD}
\end{figure}

The UQ is crucial for assuring the results produced by mathematical modeling in engineering applications. The standard deviation or variance can be considered a safety bound or a confidence interval around the mean values. Hence, we apply the methods discussed in the previous sections to an engineering application in this section. We analyze the non-ideal supersonic compressible flow within a 2D converging-diverging nozzle shown in Figure \ref{PD} \cite{WinNT}. The nozzle is 0.123 m long, with a throat height of 0.0084 m and an inlet height of 0.036 m \cite{WinNT}. This is a test case provided in SU2 for deterministic CFD simulations \cite{WinNT}. It has a simple geometry where the flow accelerates from subsonic to supersonic speeds. It can be used to investigate compressible flows where simple ideal gas laws are not enough to describe thermodynamic behavior properly. To determine the performance of the nozzle over a wide range of inlet conditions, CFD simulations are first used to confirm its performance. Then further CFD simulations are combined with uncertainty quantification methods. Additionally, the uncertainty bounds for the nozzle performance in terms of pressure and Mach number, flow density, and temperature fields along the centerline are evaluated with respect to input uncertainties. They are shown with mean and standard deviation values.\\
 
 \subsection{Test Case Description}

\begin{figure}[htbp]
 \centering
 \includegraphics[width=0.8\textwidth]{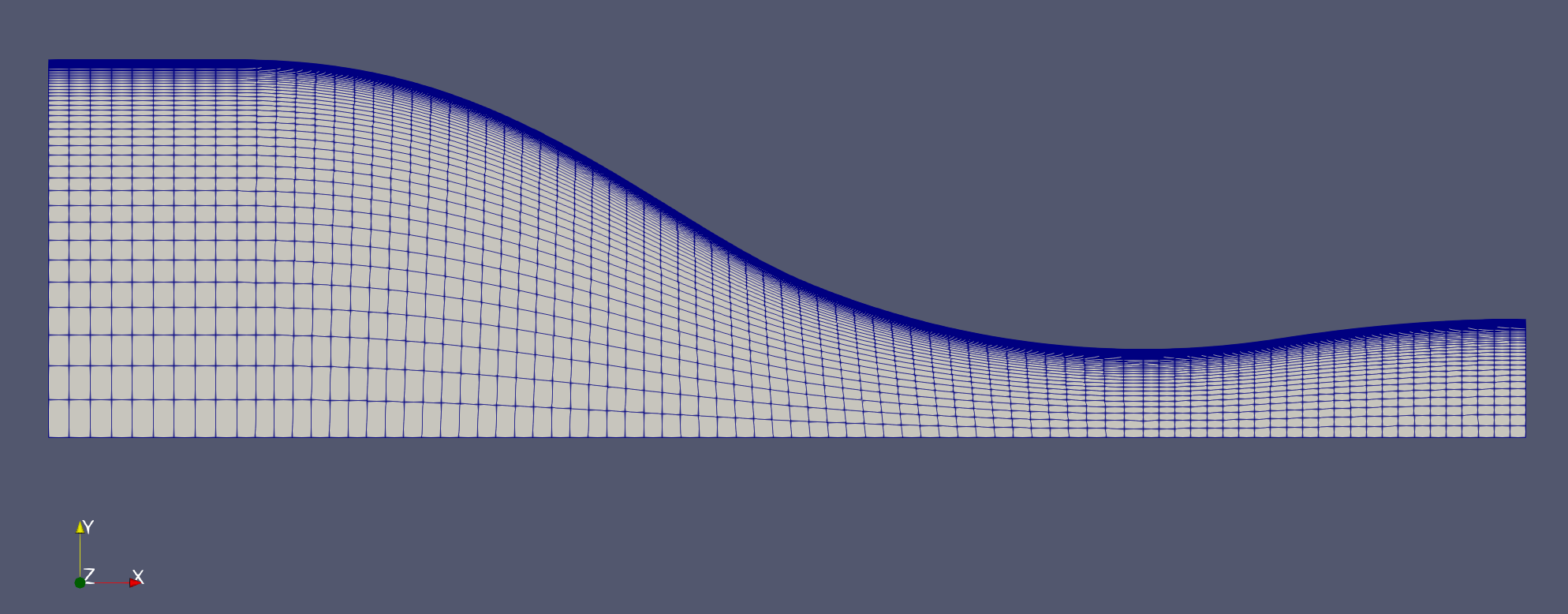}\\
 \caption{Computational domain and mesh}\label{PD1}
\end{figure}

Octamethyltrisiloxane (MDM), a pressure-sensitive fluid, is used as a working fluid for analyzing non-ideal supersonic compressible flow inside a converging-diverging (CD) nozzle. Table \ref{tabCFD} presents details about the properties of fluid and flow conditions. This configuration results in a total exhaust pressure ratio of 3.125, which results in a supersonic outflow at Mach number 1.5 \cite{WinNT}. The static pressure applied to the test case's outlet is 200,000 Pa. The computational domain and mesh are depicted in Figure \ref{PD1}. The mesh is composed of 3,540 quadrilateral elements and 3,660 nodes \cite{WinNT}. At the inlet and outlet boundaries, Riemann boundary conditions based on characteristics are used. Symmetry boundary conditions define symmetry boundaries. By mirroring the flow around the x-axis, the mesh size is reduced, along with the computational cost. On the boundary of a wall, Navier-Stokes adiabatic wall conditions are applied.\\

\begin{table}[htbp]
 \centering
 \caption{Flow conditions for C-D Nozzle CFD simulation}
 \begin{tabular}{|l|r|r|r|}
 \hline
 Parameters & \multicolumn{1}{l|}{Values} \\
 \hline
 Working fluid & Octamethyltrisiloxane \\
 \hline
Inlet Pressure & 904388 Pa \\
 \hline
Inlet Temperature &542.13 K \\
 \hline
Turbulence model & SST \\
 \hline
Gamma Value &1.01767 \\
 \hline
Gas Constant & 35.17 \\
 \hline
Critical Temperature & 565.3609 \\
 \hline
Critical Pressure & 1437500 \\
 \hline
$\mu$ & 1.21409E-05 \\
 \hline
$K_{T}$ & 0.030542828 \\
 \hline
 \end{tabular}%
 \label{tabCFD}%
\end{table}%

\subsection{Deterministic Results}

For deterministic simulations, the SU2 solver is run at the fixed boundary conditions, flow conditions, and fluid properties as described earlier. The total number of iteration was given 1000 so that all solutions and residuals are converged nicely. We post-process the data with Paraview (a multi-platform, open-source data analysis and visualization application). In Figure \ref{PD2}, solution fields for pressure, temperature, Mach number, and flow density are exhibited for the whole computational domain. Further, these quantities are also shown at the centerline of the nozzle for better understanding and analysis. At the inlet of the nozzle, pressure, temperature, and density of the fluid are at their maximum and then at their minimum near the outlet. In an inlet, the Mach number can be viewed as a minimum, and at the exit, the Mach number reaches a maximum value. \\

 \begin{figure}[htbp]
 \centering
 \includegraphics[width=0.6\textwidth]{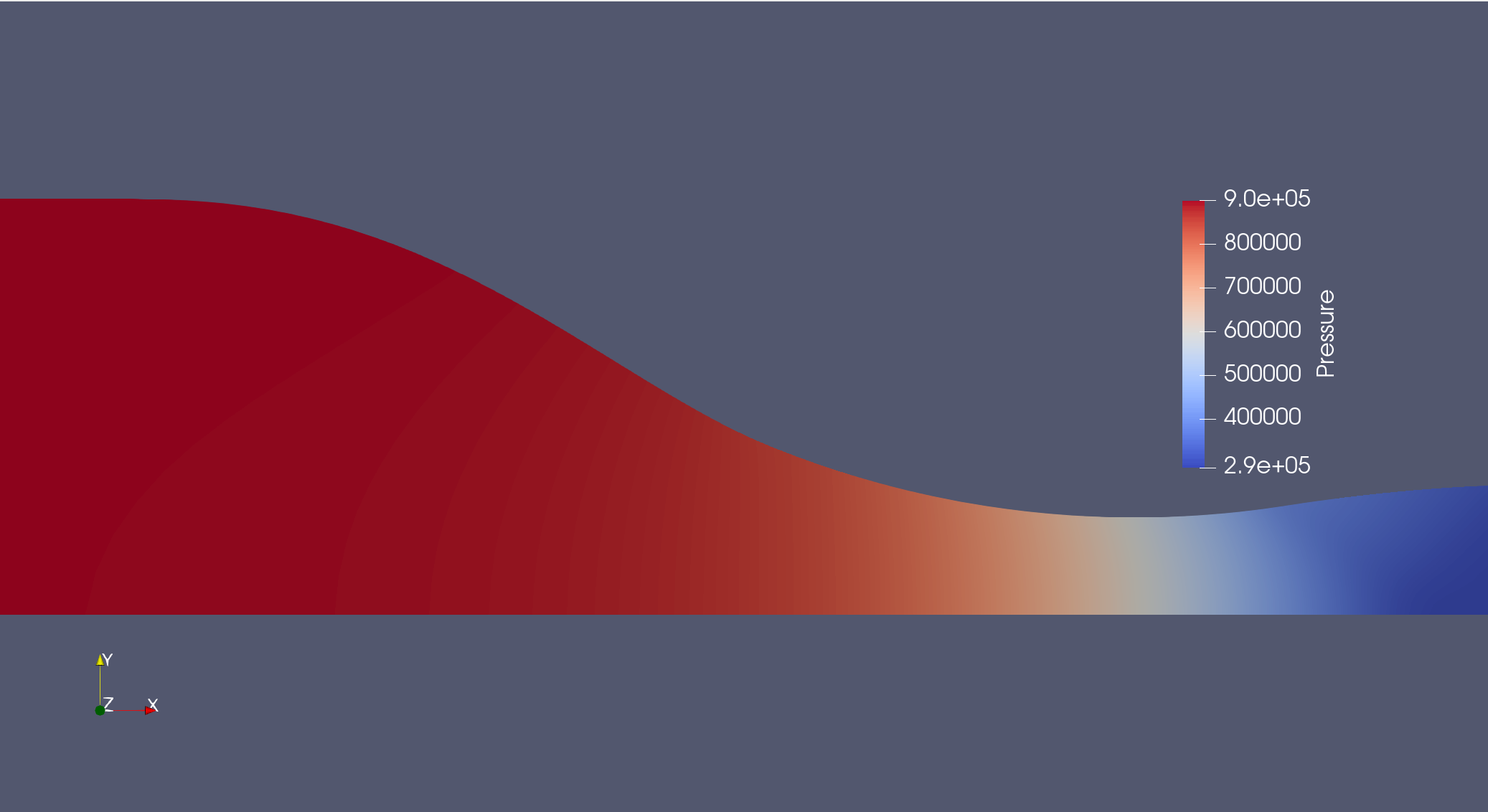}
 \includegraphics[width=0.6\textwidth]{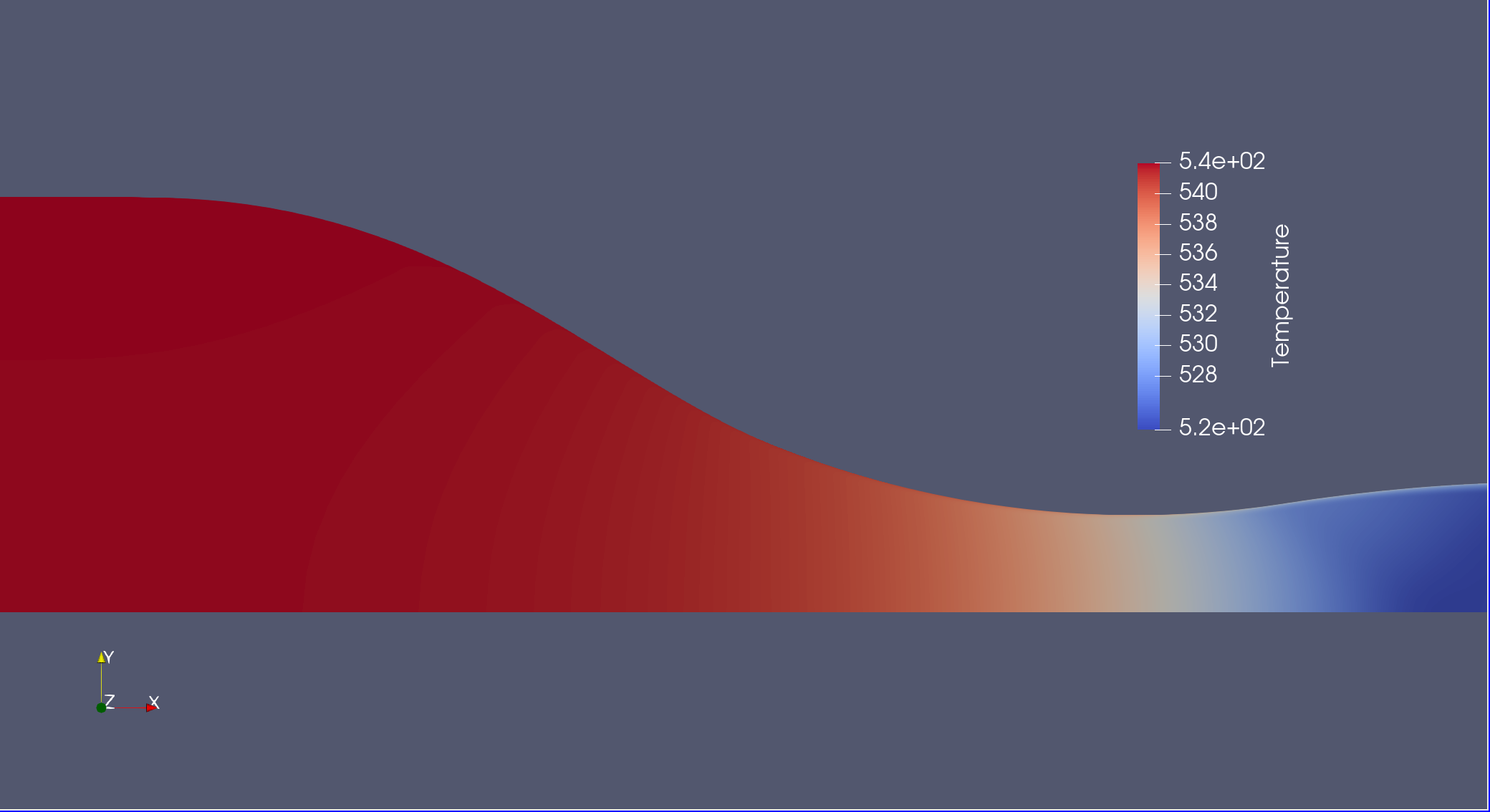}
 \includegraphics[width=0.6\textwidth]{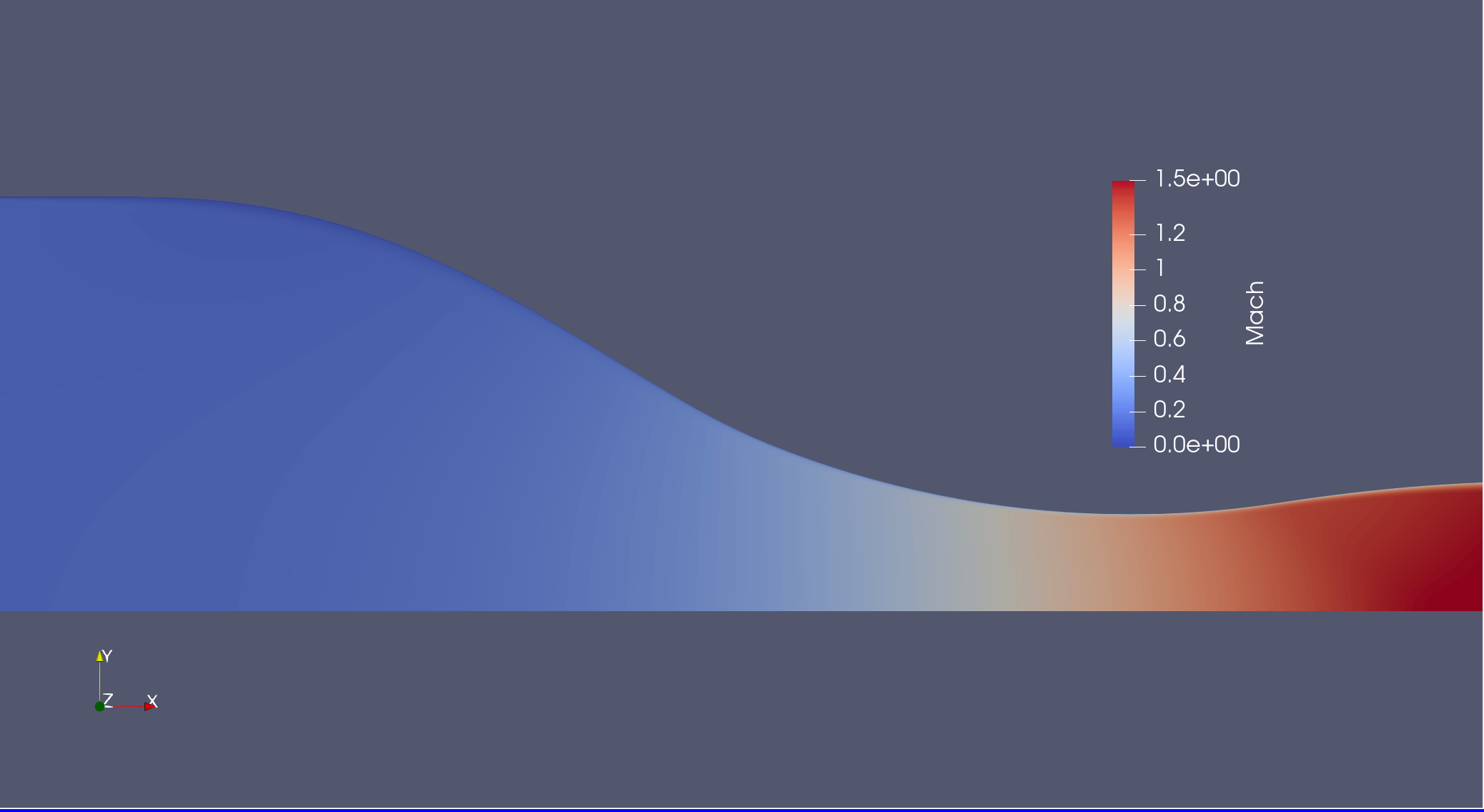}
 \includegraphics[width=0.6\textwidth]{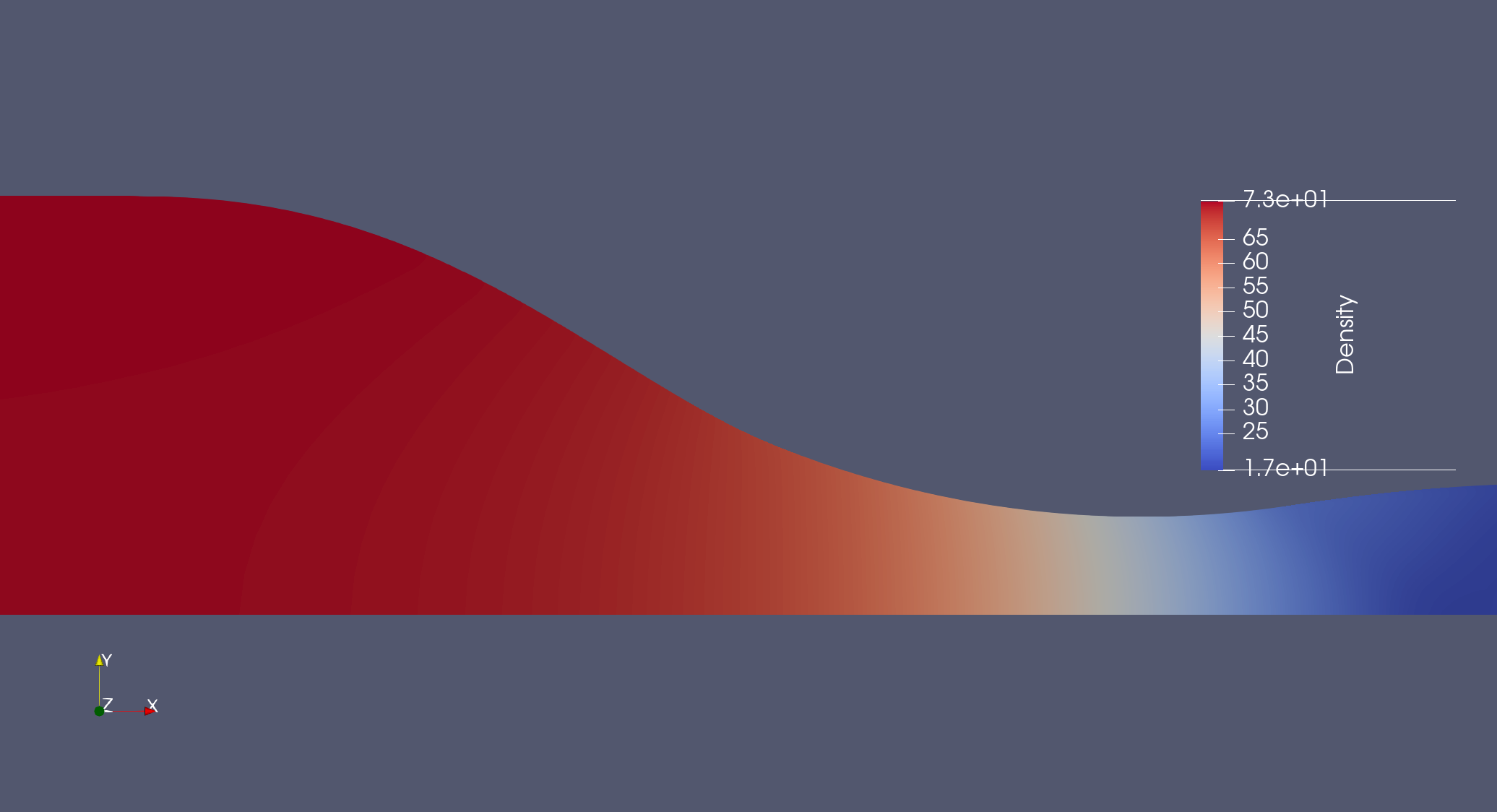}
 \caption{Numerical results (provided serially in vertical format): pressure, temperature, Mach, and density fields}\label{PD2}
\end{figure}

 \begin{figure}[htbp]
 \centering
 % Requires \usepackage{gr1phicx}
 \includegraphics[width=1.0\textwidth]{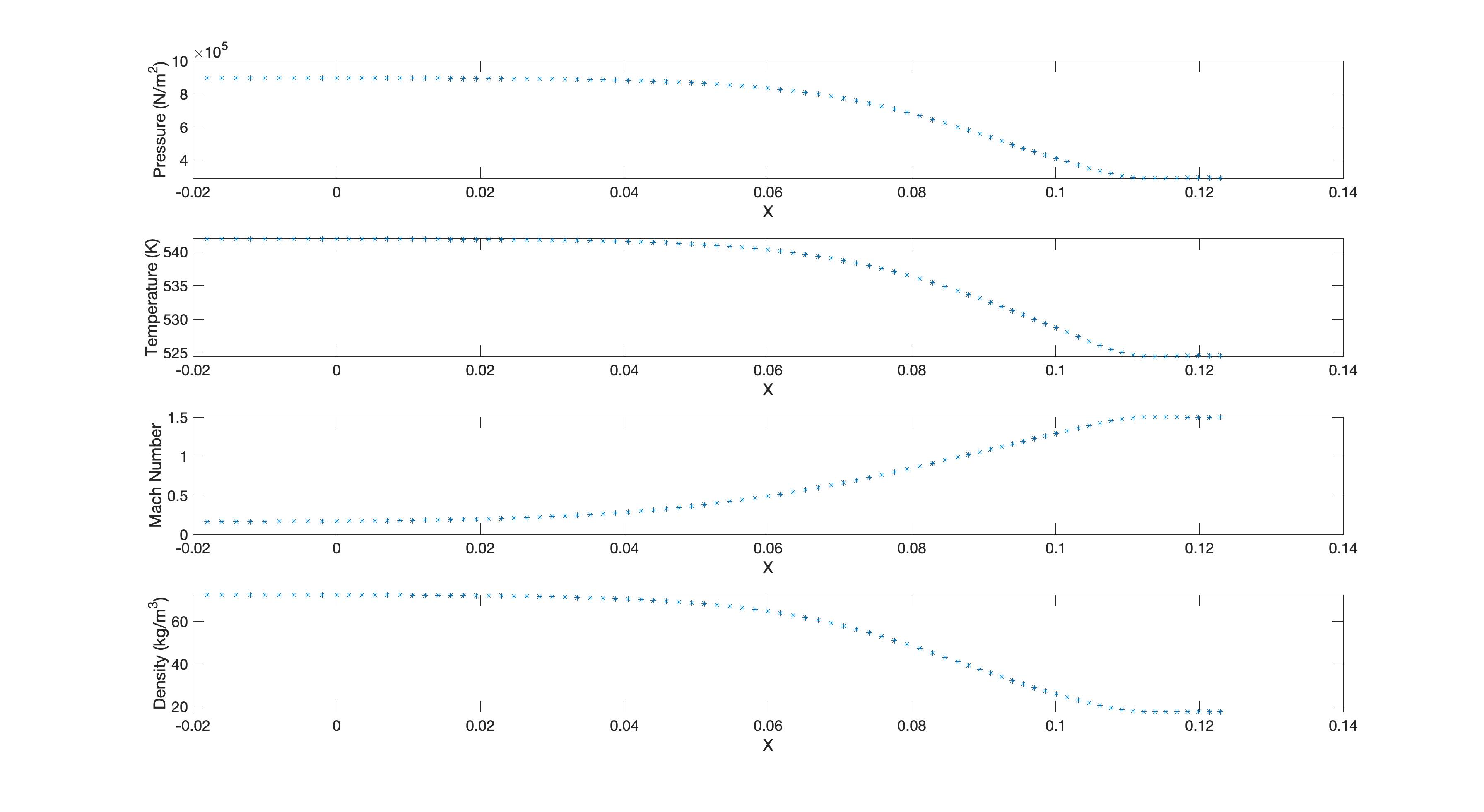}\\
 \caption{Computational results: pressure, temperature, Mach number and fluid density at the centerline of the nozzle )}\label{PD2}
\end{figure}

 \begin{figure}[htbp]
 \centering
 \includegraphics[width=1.0\textwidth]{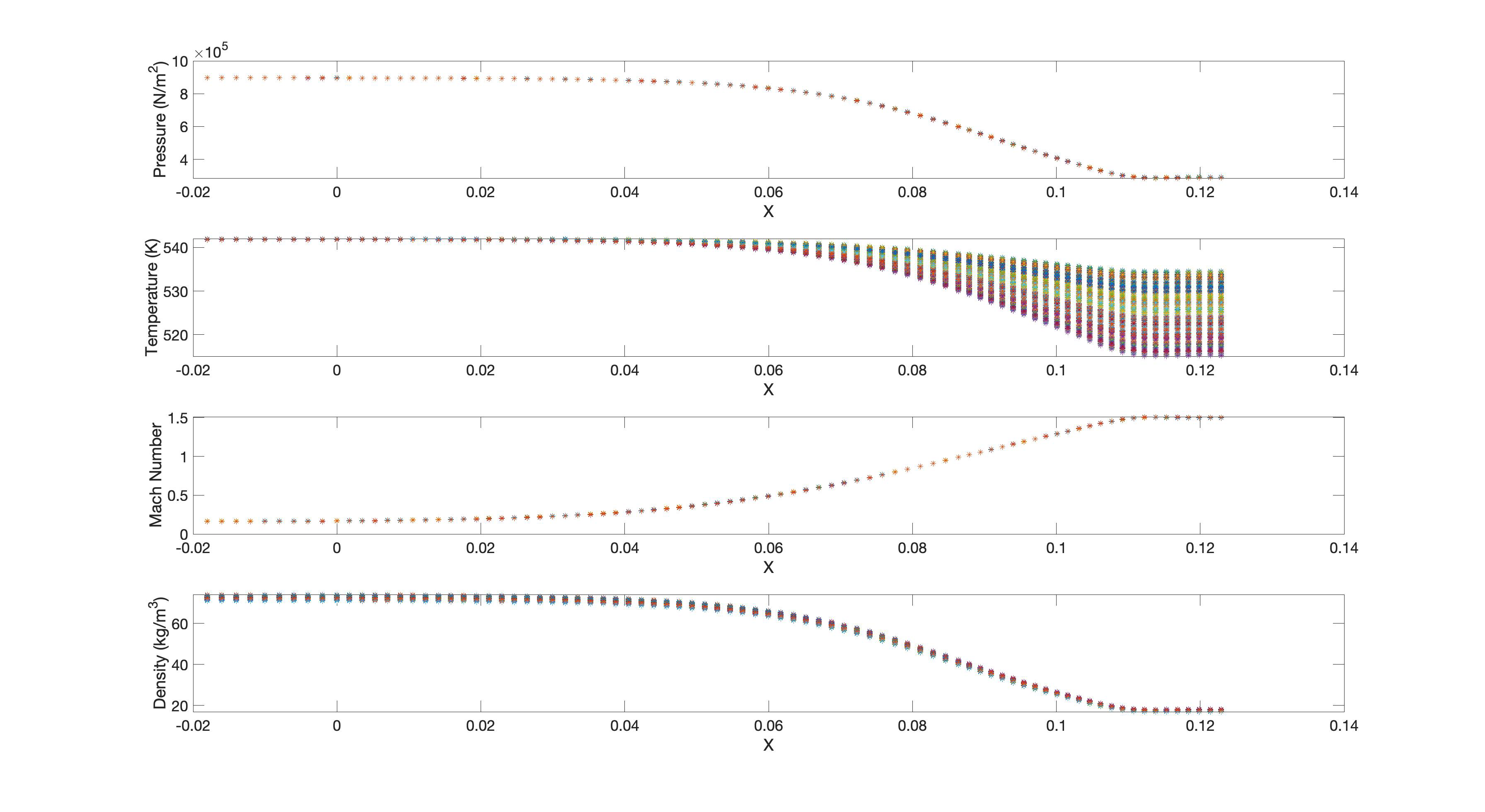}\\
 \caption{Solution samples for Pressure, Temperature, Mach number and fluid density at the centerline of the nozzle )}\label{PD3}
\end{figure}

 \begin{figure}[htbp]
 \centering
 \includegraphics[width=1.0\textwidth]{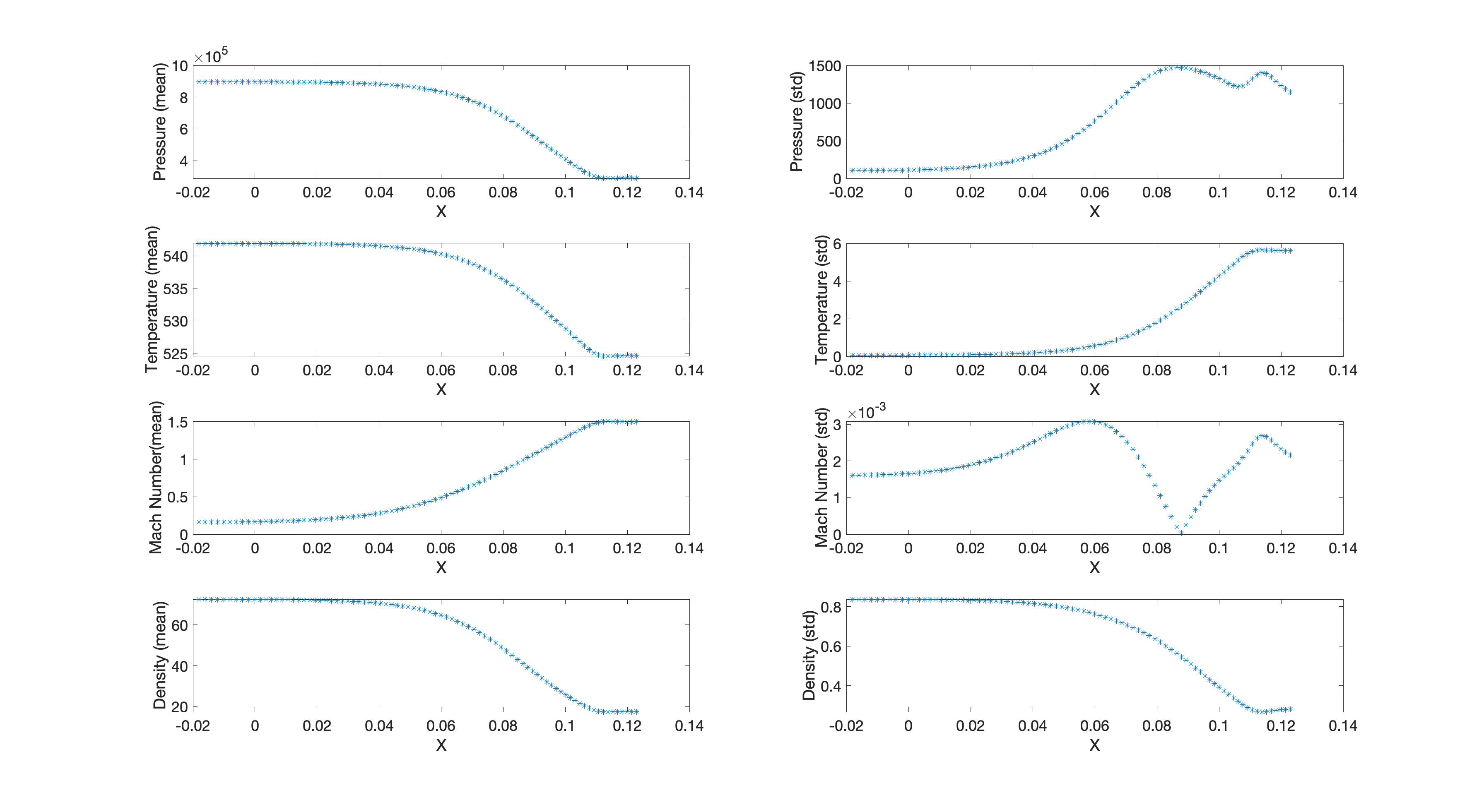}\\
 \caption{Mean and standard deviation for Pressure, Temperature, Mach number and fluid density at the centerline of the nozzle )}\label{PD4}
\end{figure}

\subsection{Description of Uncertainties}

For the uncertainty analysis, seven input parameters; two from boundary values (inlet temperature and inlet pressure), three from gas properties (specific heat ratio $\gamma$, gas constant $R$ and acentric factor $\omega$), and two from the viscosity model (molecular viscosity $\mu$ and molecular thermal conductivity $K_T$) are considered as uncertain. All parameters are considered uniformly distributed. The inlet pressure and acentric factor are assumed to have $5\%$ variability from the mean value. Gas constant, molecular viscosity, and molecular thermal conductivity are assumed to vary $2\%$ from their mean values. Minor uncertainties of $1\%$ from their mean values are given to the inlet temperature and Gamma values. All the mean values for these parameters, their uncertainties, and their ranges of variability are described in Table \ref{tabCFDUQ}.\\

\begin{table}[htbp]
 \centering
 \caption{Input uncertainties for CFD simulations}
 \begin{tabular}{|l|r|r|r|r|}
 \hline
 Parameters & \multicolumn{1}{l|}{Values} & \multicolumn{1}{l|}{Uncertainties ($\%$)} & \multicolumn{1}{l|}{Minimum} & \multicolumn{1}{l|}{Maximum}\\
 \hline
Inlet Pressure (Pa) & 904388  & 5& 859168 & 949607 \\
 \hline
Inlet Temperature (K) &542.13  & 1 & 536.71 & 547.55 \\
 \hline
Gamma Value &1.01767 & 1 & 1.00749& 1.02785 \\
 \hline
Gas Constant & 35.17 & 2& 34.47 & 35.87\\
 \hline
$\mu$ & 1.21409E-05& 2& 1.18981 & 1.23837 \\
 \hline
$K_{T}$ & 0.030542828 &2 &  0.029931971 & 0.031153684 \\
 \hline
Acentric factor ($\omega$) & 0.524 & 5& 0.498 &  0.550\\
 \hline
 \end{tabular}%
 \label{tabCFDUQ}%
\end{table}%

\subsection{Uncertainty Analysis}

As described in the previous section, the PC-Kriging method is used here to estimate the combined impact of all input uncertainties on the system responses of the CD nozzle. Usually, in the regression-based polynomial chaos method, a total of 240 CFD samples (for PC order 3 and 7 input uncertainties) will be required to estimate the statistical quantities (mean and standard deviance) of the output accurately (see \cite{kumar2016efficient}). Here to construct the PC-Kriging-based surrogate model, only 100 CFD simulations are used. For input parameters, 100 designs of experiments are constructed using the Sobol sequence-based sampling technique. In Figure \ref{PD3}, the CFD solutions for pressure, temperature, Mach number, and fluid density along the nozzle centerline are shown for all 100 samples. It can be seen that pressure and Mach number are not varying much with the input uncertainties. However, minor variations can be seen in density with the input variations. The most significant variations can be seen for the temperature field. In Figure \ref{PD4}, the mean and standard deviation are shown for all these quantities. These values are calculated from the PC-Kriging based surrogate model. The mean values behavior is similar to the deterministic solutions. For pressure, temperature, and Mach number, the standard deviation values are higher at the outlet. That means the highest fluctuations are at the nozzle outlet. However, the standard deviation for fluid density is seen lower at the nozzle outlet.\\

It is also important to address that this developed uncertainty method can be applied to other domains such as nuclear engineering in terms of safety assessment of advanced reactor system \cite{kumar2021quantitative}. In addition, the authors also utilized this methodology to understand the evaluate the uncertainties in composite materials \cite{kumar2021multi}.

\section{Conclusions}
In this work, first, we describe the two most popular meta-modeling methods (Polynomial Chaos and Kriging methods) suitable for uncertainty quantification in engineering applications. Further, to increase the efficiency, the polynomial chaos and Kriging methods are combined and used for an engineering test problem under multiple uncertainties. A 2D supersonic converging-diverging nozzle is considered for the analysis where the multi-physics CFD solver SU2 is used for deterministic solutions. The UQ methods (polynomial chaos, Kriging, and PC-Kriging) are developed in Matlab and are further combined with SU2 for uncertainty quantification. The standard deviation can be considered as a safety bound or a confidence interval around the mean values. Hence, for assurance in making crucial decisions, the results are discussed in terms of the mean and standard deviation of the output quantities, i.e., pressure, temperature, Mach number, and fluid density.\\

Future work will focus on its application in multiscale modeling of composite accident-tolerant nuclear fuels with Sic/Sic claddings for small modular reactor (SMR) applications.

\section*{Acknowledgement}
The computational part of this work was supported in part by the National Science Foundation (NSF) under Grant No. OAC-1919789.

\bibliographystyle{unsrtnat}
\bibliography{references}  %%% Uncomment this line and comment out the ``thebibliography'' section below to use the external .bib file (using bibtex) .





% \end{thebibliography}

\end{document}